# Multi-Frequency Reverberant Shear Waves for Assessing Tissue Dispersion in Optical Coherence Elastography

HAMIDREZA ASEMANI,[1,2,*] PANOMSAK MEEMON,[3,1] GILMER FLORES BARRERA,[4] JANNICK P. ROLLAND,[1,4] AND KEVIN J. PARKER[2,4]

[1]Institute of Optics, University of Rochester, Rochester, New York 14627, USA
[2]Department of Electrical and Computer Engineering, University of Rochester, Rochester, New York 14627, USA
[3]School of Physics, Institute of Science, Suranaree University of Technology, Nakhon Ratchasima 30000, Thailand
[4]Department of Biomedical Engineering, University of Rochester, Rochester, New York 14627, USA
*hasemani@ur.rochester.edu

**Abstract:** Optical coherence elastography (OCE) is a powerful non-invasive imaging technique for high-resolution assessment of tissue elasticity and viscoelasticity. Accurate characterization of viscoelastic properties requires estimating shear wave speed (SWS) across multiple frequencies, as dispersion induces frequency-dependent variations in wave speed. This study introduces a novel multi-frequency reverberant OCE (MFR-OCE) approach to enhance viscoelastic tissue characterization by simultaneously capturing shear wave dynamics over multiple frequencies. We present the theoretical framework, experimental setup, and validation of MFR-OCE through simulations and experiments on gelatin phantoms, *ex vivo* porcine cornea, and *ex vivo* bovine liver. Simulation results demonstrate that MFR-OCE estimates SWS with errors below 4% relative to ground truth, and phantom experimental results show that MFR-OCE and single-frequency OCE also yield closely matching SWS estimates, with differences below 3%. Furthermore, the frequency-dependent dispersion coefficients extracted from biological tissues and phantoms align with the theoretical viscoelastic power law model. The gelatin phantoms exhibit a low viscoelastic behavior with an exponent of 0.13 for the power law fit of SWS, while the *ex vivo* porcine cornea demonstrates intermediate viscoelastic behavior, with a power law exponent of 0.33. The liver tissue shows significant frequency dependence, with a power law exponent of 0.51. These findings demonstrate that MFR-OCE enables a more comprehensive understanding of tissue mechanics and holds the potential for improving diagnostic accuracy in clinical applications.

## 1. Introduction

In the realm of biomedical imaging, the demand for non-invasive, high-resolution techniques for assessing tissue biomechanics continues to drive innovation. Optical coherence elastography (OCE) has emerged as a promising technique, offering high-resolution, non-invasive imaging for evaluating tissue mechanical properties *in vivo* [1-3]. Among various OCE techniques, shear wave-based methods have gained substantial attention due to their ability to generate quantitative elasticity maps [4,5] and provide detailed insights into tissue viscoelastic properties [6,7]. Compared to other elastography modalities, such as magnetic resonance elastography and ultrasound elastography, which typically offer millimeter-scale imaging resolution [8,9], OCE is distinguished by its ability to achieve micrometer-scale (∼2–10 $\mu$m) imaging resolution [10-12].

Numerous studies have explored the elastic properties of biological tissues using wave-based OCE. These investigations include measurements of the shear modulus and Young's modulus of the cornea [13-16], Young's modulus of the porcine liver [17], and the elasticity of mouse brain tissue [18-20]. In purely elastic materials, waves propagate without energy loss, resulting in a single, consistent velocity across all frequencies. However, in viscoelastic materials, loss mechanisms cause frequency-dependent variations in wave speed, a phenomenon known as dispersion [21,22]. The dispersion curves and viscoelastic properties of biological tissues have been examined in various studies, including the dispersion curve of porcine cornea [23-25], viscoelastic characterization of porcine cornea [26], and shear viscosity of chicken liver [27]. Additionally, Poul *et al.* [28] demonstrated that the viscoelastic behavior of bovine liver tissue can be modeled using a two-parameter power-law model for shear wave speed (SWS) dispersion.

Reverberant shear wave fields provide a novel framework for enhancing the characterization of tissue biomechanics in elastography [29-31]. These fields arise from the superposition of shear waves propagating in multiple directions, including reflections from tissue boundaries and internal heterogeneities. Recent advancements have highlighted the potential of this method for probing complex biological tissues, where conventional single-point techniques may fail to capture the full mechanical behavior [32-34].

To accurately analyze the viscoelastic and lossy nature of the biological tissues using shear wave elastography, it is essential to estimate SWS across multiple frequencies [35,36]. However, several shear wave OCE methods rely on a single excitation frequency, limiting their ability to fully capture complex tissue mechanics, particularly the dispersive viscoelastic properties. Other impulsive excitation methods can be used to impart a band of frequencies within a propagating transient wave [10]; however these decay rapidly with distance away from the source location.

To address this limitation, we introduce a novel OCE approach based on multi-frequency reverberant shear waves (MFR-OCE). The advantages of this approach include robust shear wave propagations within a 3D volume and simultaneously across a discrete set of frequencies covering a wide bandwidth. This paper presents the theoretical framework, simulations, experimental setup, and preliminary results of MFR-OCE on gelatin phantoms, porcine cornea, and bovine liver, demonstrating its potential to enhance both elastic and viscoelastic tissue characterization for clinical applications.

## 2. Mathematical concepts

The particle velocity within a fully reverberant shear wave field is mathematically represented as [37]

$$\boldsymbol{V}(\boldsymbol{\varepsilon}, t) = \sum_{q,l} \widehat{\boldsymbol{n}}_{ql}\, v_{ql}\, e^{i(k\widehat{\boldsymbol{n}}_q \cdot \boldsymbol{\varepsilon} - \omega_0 t)}, \tag{1}$$

where $\widehat{\boldsymbol{n}}_q$ represents the direction of wave propagation, with the index $q$ denoting a specific instance of the random unit vector $\widehat{\boldsymbol{n}}_q$. The vector $\widehat{\boldsymbol{n}}_{ql}$ indicates the direction of particle velocity for that instance of $q$, and the index $l$ identifies a particular instance of the random unit vector $\widehat{\boldsymbol{n}}_{ql}$. $v_{ql}$ is an independent, identically distributed random variable representing the magnitude of particle velocity for that instance of $q$. In the context of the shear wave propagation, $\hat{n}_q$ and $\hat{n}_{ql}$ are perpendicular, resulting in $\hat{n}_q \cdot \hat{n}_{ql} = 0$. In this expression, $k$ represents the wavenumber, $\omega_0$ denotes angular frequency, $\boldsymbol{\varepsilon}$ is the position vector, and $t$ is time.

In OCE, the particle velocity is typically measured along the laser axis, which is perpendicular to the sample surface. Assuming the sensor axis aligns with the *z*-axis, the recorded particle velocity can be expressed as $V_z(\boldsymbol{\varepsilon}, t) = \boldsymbol{V}(\boldsymbol{\varepsilon}, t) \cdot \hat{e}_z$, where $\hat{e}_z$ is a unit vector in the *z*-direction. For the reverberant field $V_z(\boldsymbol{\varepsilon}, t_0)$, closed-form complex analytical solutions can be derived using spatial autocorrelation in the spherical coordinate system as detailed by Aleman-Castañeda *et al.* [38] as

$$B_{V_z V_z}(\Delta \varepsilon) = 3\overline{V_z}^2 \left\{ \frac{\sin^2 \theta_s}{2} \left[ j_0(k\Delta\varepsilon) - \frac{j_1(k\Delta\varepsilon)}{k\Delta\varepsilon} \right] + \cos^2 \theta_s \frac{j_1(k\Delta\varepsilon)}{k\Delta\varepsilon} \right\} \quad (2)$$

where $B_{V_z V_z}$ represents the autocorrelation function of $V_z$, while $\overline{V_z}^2$ denotes the expected value of the squared particle velocity magnitude $v_{ql}^2$ averaged over both $q$ and $l$ instances. The functions $j_0$ and $j_1$ are spherical Bessel functions of the first kind of zero and first order, respectively, and $\theta_s$ denotes the angle between $\Delta\varepsilon$ and the $z$-axis. By performing an angular integration of the autocorrelation function over $\theta_s$ from 0 to $2\pi$ within two-dimensional planes, the angular integral autocorrelation ($B_{AIA}$) expressions for the $xy$, $xz$, and $yz$ planes are derived [39] as

$$B_{AIA_{xy}}(\Delta\rho) = \frac{3}{2} \overline{V_z}^2 \left[ j_0(k\Delta\rho) - \frac{j_1(k\Delta\rho)}{k\Delta\rho} \right] \quad (3.a)$$

$$B_{AIA_{xz}}(\Delta\rho) = B_{AI_{yz}}(\Delta\rho) = \frac{3}{4} \overline{V_z}^2 \left[ j_0(k\Delta\rho) + \frac{j_1(k\Delta\rho)}{k\Delta\rho} \right] \quad (3.b)$$

where $\Delta\rho$ represents the one-dimensional lag in the angular integral autocorrelation argument. The local wavenumber $k$ is extracted by analyzing the 2D autocorrelation function within a localized region of a reverberant field and fitting the resulting autocorrelation profiles to Equation 3.a or 3.b, depending on the plane configuration. Given the excitation frequency $\omega$, the shear wave speed $C_s$ is determined using the relation $C_s = \omega/k$.

To implement the angular integral autocorrelation approach in a multi-frequency reverberant shear wave field, it is essential to extract the particle velocity at each frequency using a bandpass filter. This method enables the separate estimation of SWS for each frequency. By incorporating multiple frequencies into the SWS estimation process, a more comprehensive analysis of shear wave behavior is achieved, leading to improved diagnostic capabilities and deeper insight into tissue dynamics.

The power-law model provides a mathematical framework for characterizing the frequency-dependent dispersion of SWS in viscoelastic bio-materials. Within the Kelvin-Voigt fractional derivative model, this relationship is expressed as

$$C_s(f) = C_0 f^{\frac{a}{2}} \quad (4)$$

where $f$ is the frequency, $C_0$ represents the reference wave speed at a unit reference frequency (e.g., 1 Hz), and $a$ is the dispersion coefficient [28] of the complex shear modulus.

## 3. Multi-frequency reverberant elastography simulation

The k-Wave toolbox in MATLAB (version 2022b, The MathWorks, Inc., Natick, MA, USA) [40] was employed to assess the effectiveness of multi-frequency reverberant shear wave fields for SWS estimation. The simulation involved a two-sided medium, consisting of a softer side with an SWS of 1 m/s and a stiffer side with an SWS of 2 m/s, as depicted in Fig. 1(a). The model was constructed as a cube with dimensions of 12 mm × 12 mm × 2.4 mm. Both the softer and stiffer sides were modeled as homogeneous isotropic materials with a density of 1000 kg/m³.

In order to generate a multi-frequency reverberant shear wave field, 1000 point-velocity excitation sources were utilized. The region of interest (ROI) was defined as a smaller cube with dimensions of 10 mm × 10 mm × 2 mm (Fig. 1). To create a source-free reverberant interior within the medium, the excitation sources were placed outside the ROI at random locations near the boundaries. An excitation signal comprising five frequencies (i.e., 500 Hz, 1000 Hz, 1500 Hz, 2000 Hz, and 2500 Hz) with randomly assigned amplitude and phase was applied to these point sources. The wave equations were solved using a time interval of 5 $\mu$s over 2667 time steps, allowing 13.3 ms for wave propagation to reach a steady state and the formation of a fully reverberant shear wave field. White Gaussian noise with a signal-to-noise ratio of 10 dB was added to the shear-wave field. Figure 1(b) illustrates the 3D particle velocity field of the multi-frequency reverberant shear wave within the two-sided medium, highlighting wavelength variations on each side. The simulation was designed to mimic real OCE experiments, with the velocity field analyzed only along the sensor axis (*z*-axis). To estimate the SWS at different frequencies, the velocity field for each frequency was extracted from the multi-frequency reverberant shear wave field using bandpass frequency filters with a range of 10 Hz centered at 500 Hz, 1000 Hz, 1500 Hz, 2000 Hz, and 2500 Hz.

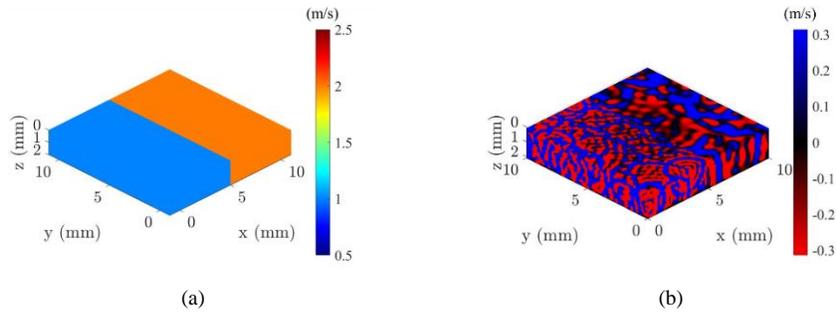

Fig. 1. K-Wave elastography simulation. (a) A two-sided medium with a softer side (SWS of 1m/s) and stiffer side (SWS of 2m/s); (b) 3D particle velocity field of the multi-frequency reverberant shear wave field including excitation frequencies of 500 Hz, 1000 Hz, 1500 Hz, 2000 Hz, and 2500 Hz.

## 4. Multi-frequency reverberant OCE

### 4.1 Experimental setup

Figure 2 illustrates the schematic of the MFR-OCE system, which integrates a custom-built, fiber-based, swept-source OCT system with a piezoelectric actuator-based excitation mechanism. The swept-laser source (HSL-2100-HW, Santec, Aichi, Japan) has a spectral tuning range of approximately 140 nm, centered at a wavelength of 1310 nm. The system includes a Mach-Zehnder interferometer (MZI) for spectral nonlinearity calibration, as described by Yao *et al.* [41,42]. The measured depth resolution in air is approximately 6 $\mu$m.

The sample arm incorporates a custom pupil-relay precision scanning mechanism in the sample arm to achieve diffraction-limited performance, as detailed by Xu *et al.* [43]. A Thorlabs microscope objective (model LSM03, Newton, NJ, USA) with an effective focal length of 36 mm and a maximum field of view of $18 \times 18$ mm$^2$ is used for scanning. The system's lateral resolution is measured to be approximately 20 $\mu$m.

The system employs two balanced photodetectors (model 1817-FC, Newport, Irvine, California, USA) to capture the two primary signals: (1) the interference signal between the sample and reference arms and (2) the MZI calibration signal. These detectors feature a bandwidth ranging from DC to 80 MHz and are sensitive to wavelengths between 900 nm and 1700 nm. To maximize fringe contrast, two polarization controllers (Thorlabs FPC030) are used, i.e., one in the reference arm and one in the sample arm, for the OCT signal. Similarly, an additional polarization controller is used to maximize the fringe contrast of the MZI signal. The system sensitivity is measured to be approximately 110 dB.

System operation is controlled using LabVIEW software (Version 14, National Instruments, Austin, TX, USA), which allows for adjustment of parameters such as the field of view, the number of lateral scanning points (B-mode scan), and the number of spectra to acquire at each lateral position (M-mode scan). A high precision function generator (AFG3021C, Tektronix, Beaverton, OR, USA) is used to synchronize the timing between the mechanical excitation and the OCT acquisition.

The excitation system consists of a function generator (model 4052, B&K Precision, Yorba Linda, CA, USA) and an amplifier (PDu150, PiezoDrive, Callaghan, New South Wales, Australia). A multi-frequency signal is generated in MATLAB, imported into the excitation system, and used to generate a synchronized multi-frequency excitation signal. This signal is then applied to a piezoelectric actuator, which induces multi-frequency shear waves within the sample.

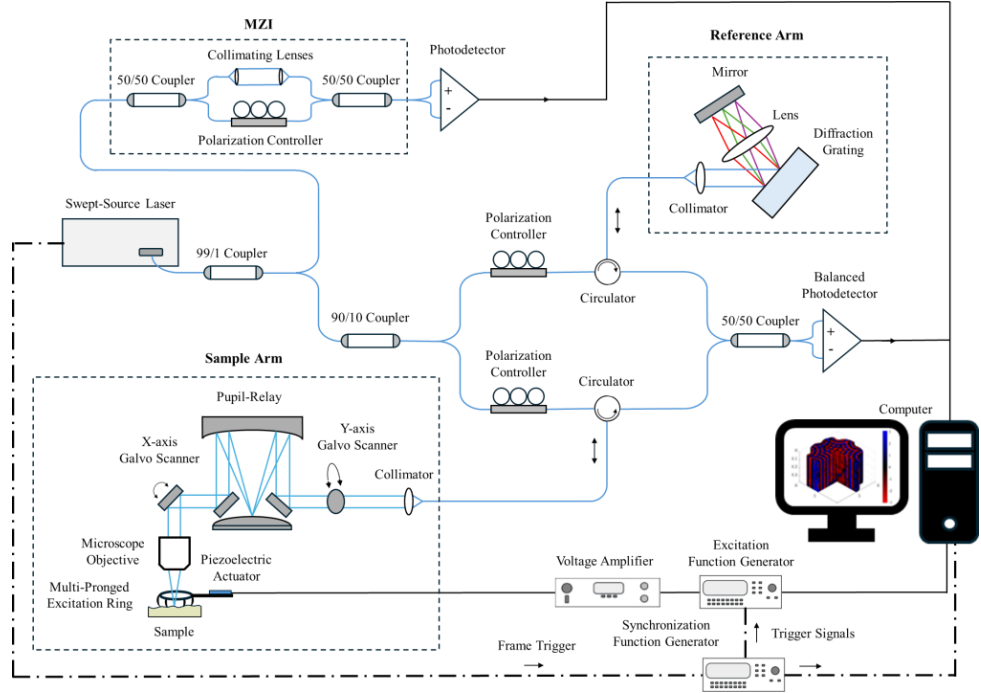

Fig. 2. Detailed schematic of the MFR-OCE system.

*4.2 Multi-frequency signal and synchronization process*

Synchronization is a critical aspect of multi-frequency shear wave elastography as it ensures that the OCT imaging system is synchronized not only with the main excitation signal but also with each individual frequency component. In conventional swept-source OCT, the 3D data acquisition process involves a combination of 2D lateral scanning of the laser beam on a sample's surface within the field of view (B-mode scanning), and depth scanning along the beam propagation direction (axial scan), which is achieved by the sweeping of the laser wavelength. The acquired data is then used to form a 3D tomographic dataset.

In MFR-OCE, each point on the 2D surface is scanned multiple times (multiple sweeps of the laser wavelength to capture sample motion, a so-called "M-mode scanning". As the waves propagate during the scan, the axial position of tissue particles changes, introducing a Doppler phase shift across the time sampling. Therefore, analyzing the Doppler phase shift on the time-series data acquired through M-mode scanning enables the estimation of SWS. In this study, 100 spectra per M-mode scan and 100 × 100 lateral sampling points per B-mode scan were acquired and processed.

Figure 3 illustrates the synchronization process in MFR-OCE scanning. The top axes represent B-mode scanning for spatial data acquisition, the middle axes depict mechanical excitation, and the bottom axes show M-mode scanning at each position. At the end of the scan, a 4D dataset is generated, encompassing three spatial dimensions and time. To achieve a consistent and accurate 4D scan of the shear waves, it is essential that the wave field remains identical at each scan position on the 2D surface. This uniformity ensures that the OCT scans can be considered a comprehensive scan, allowing the acquired data to be integrated into a 4D representation of shear wave dynamics in the medium over time. In MFR-OCE, the displacement field corresponding to each excitation frequency is extracted to estimate the SWS. Thus, precise synchronization of all excitation frequencies with the imaging system is crucial for accurate multi-frequency shear wave analysis.

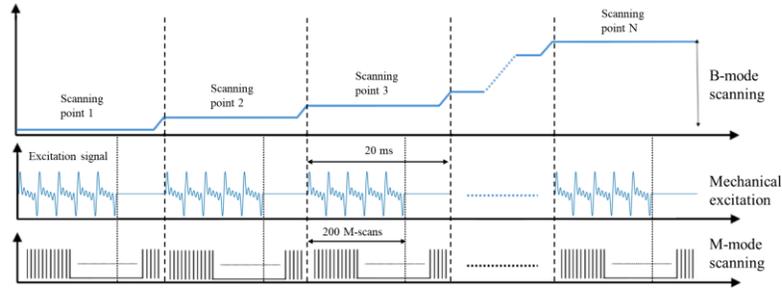

Fig. 3. Timing diagram illustrating the synchronization process between mechanical excitation and the imaging system in MFR-OCE.

As the response of the piezo actuator varies across different frequencies, the amplitude of the induced shear wave field also differs depending on the excitation frequency. To ensure robust measurements and achieve a nearly uniform amplitude across all frequencies, a tuning process was applied. The shear wave field amplitude was first estimated for each excitation frequency, and an appropriate scaling factor was then applied to the excitation signal to level wave field amplitudes across frequencies. A multi-frequency signal with five frequencies including 500 Hz, 1000 Hz, 1500 Hz, 2000 Hz, and 2500 Hz was generated in MATLAB and imported into the excitation system. Each frequency was assigned a random initial phase. Figure 4 illustrates the combination of multiple sinusoidal single-frequency waves ($V_1(t)$ to $V_n(t)$) to generate a multi-frequency excitation waveform ($V_{mf}(t)$). In the frequency domain, different excitation frequencies are distinguishable by their higher amplitudes ($A(f)$).

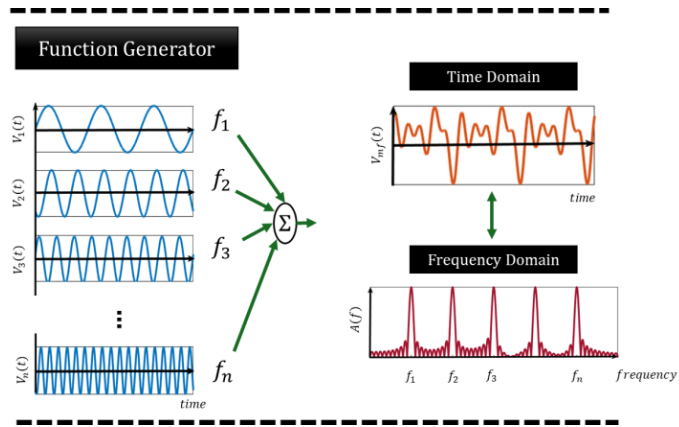

Fig. 4. Generation of a multi-frequency excitation waveform by combining sinusoidal signals with a randomly assigned initial phase.

The scanning trigger frequency was set to 50 Hz, meaning the system scans each point every 20 ms. Consequently, the excitation signal length for each scan point is also 20 ms, as illustrated by the mechanical excitation in Fig. 3. The main key to synchronization lies in the rest period within the excitation signal, where the amplitude drops to zero, allowing the imaging system sufficient time to switch scan points. The multi-frequency excitation signal was thus designed with an active period of 10 ms, followed by a 10 ms interval allocated for scan position adjustment via Galvo scanner tilting. This timing alignment ensures a stable scan for each position. The swept-source laser operates at a sweep rate of 20 kHz, resulting in an M-mode scan every 50 $\mu$s, which corresponds to 200 M-mode scans during the 10 ms period. However, only the central 100 M-mode scans are retained for elastography measurements to optimize data processing. The displacement field was obtained using the acquisition method and data postprocessing techniques developed by Zvietcovich *et al.* [34].

### *4.3 Sample preparation*

### *4.3.1 Gelatin phantom*

Four isotropic, homogeneous gelatin phantoms were prepared using standard methods [39]. Each phantom consisted of 5% gelatin powder (G1890-1KG, gelatin from porcine skin, Sigma-Aldrich, MO, USA) to provide elastic properties, 3% intralipid powder for optical scattering, 1% salt, and 91% water. The formulation and preparation method were specifically designed to ensure isotropic mechanical behavior, facilitating the investigation of multi-frequency shear wave elastography.

### *4.3.2 Cornea*

*Ex vivo* porcine whole eye globe samples were sourced from a local slaughterhouse immediately after slaughter and kept under refrigerated conditions during transport. Upon arrival at the laboratory, the eye globes were immersed in a balanced salt solution (BSS). The samples were then allowed to reach room temperature before the experiments commenced. This preparation method ensured the preservation of the cornea's natural biomechanical properties. All experiments were conducted on the day of collection, using only intact eyes with undamaged corneas. Before placement in a custom-built holder, surrounding adipose and muscular tissues were carefully removed. To maintain an intraocular pressure of 15 mmHg, a needle connected to an intravenous fluid bag containing the BSS was inserted through the holder into the eye. To prevent dehydration, the eyes were irrigated with the saline solution at regular intervals.

### *4.3.3 Liver*

A freshly harvested *ex vivo* bovine liver was obtained from a slaughterhouse immediately post-slaughter and transported under refrigerated conditions. Small samples were excised from the liver for the MFR-OCE experiments. The samples were then brought to room temperature while submerged in saline solution to prevent tissue degeneration and dehydration.

### *4.4 Generating the multi-frequency reverberant shear wave field*

To generate a reverberant shear wave field, two custom-designed multi-pronged rings with eight arms, arranged around central rings measuring 10 mm and 5 mm in diameter, were employed. The rings were gently placed on the samples to induce reverberant shear waves. In the MFR-OCE experiments on phantoms and porcine corneas, the field of view was set to 10 mm × 10 mm, with 100 scan points in each direction across the sample. A smaller field of view (5 mm × 5 mm) and a smaller multi-pronged ring (the 5 mm ring) were used in the liver experiments due to higher shear wave attenuation in liver tissue. A multi-frequency excitation signal comprising five frequencies (i.e., 500 Hz, 1000 Hz, 1500 Hz, 2000 Hz, and 2500 Hz) was applied to the samples. To compare the multi-frequency results with single-frequency results, additional experiments were conducted on phantom samples using single-frequency reverberant shear waves at 1000 Hz, 1500 Hz, and 2000 Hz.

## 5. Results and discussion

### 5.1 Multi-frequency reverberant elastography simulation

The angular integral autocorrelation (AIA) [39] method was employed to estimate the SWS across various frequencies. Given that the wavelength increases as the frequency decreases, a larger autocorrelation window size was used for lower frequencies to ensure that an adequate number of waves were included in the autocorrelation calculation. Figure 5 presents the 3D SWS maps for different excitation frequencies. The two distinct sides of the simulated medium with different SWS are clearly visible in all SWS maps. As the frequency increases, the estimated SWS becomes more uniform within each side of the medium. The variation in SWS at the boundary between the two materials is attributed to the windowing effect, an inherent limitation in this analysis.

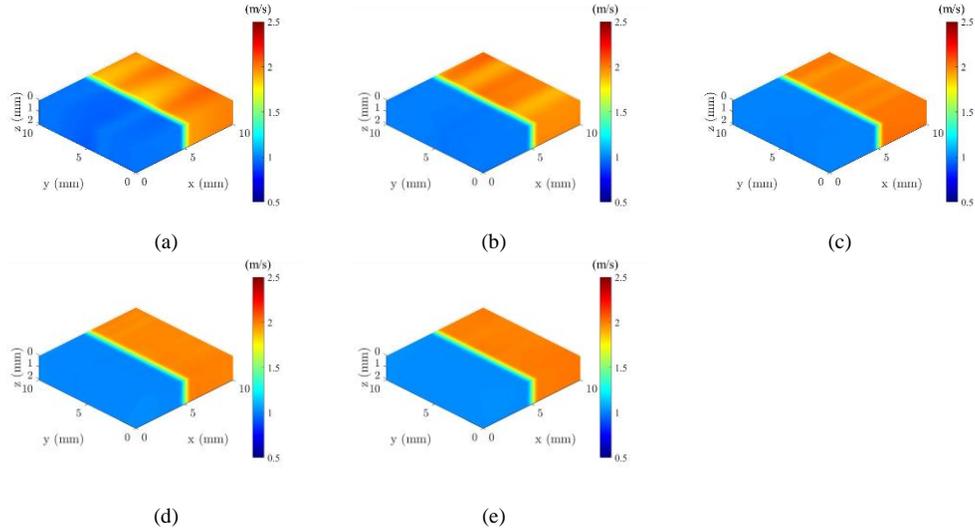

Fig. 5. Estimated SWS for the simulated two-sided (soft and hard) medium at different excitation frequencies: (a) 500 Hz, (b) 1000 Hz, (c) 1500 Hz, (d) 2000 Hz, and (e) 2500 Hz.

Figure 6 presents the mean SWS for each side of the simulated medium across different excitation frequencies. At all frequencies, the estimated SWS values on both sides closely matched the defined values for the model, with an error of less than 4%. The power law exponent was estimated to be 0.01 (higher SWS, orange) and 0.04 (lower SWS, blue), indicating a nearly dispersionless elastic response in the simulated system.

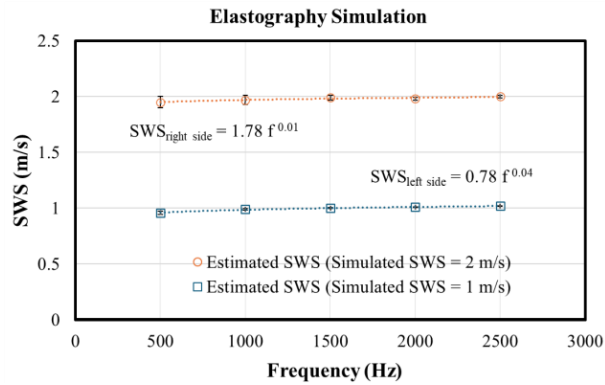

Fig. 6. The estimated SWS in the simulated two-sided medium across different excitation frequencies.

## 5.2 MFR-OCE on 5% gelatin phantom

Figure 7 presents the results for one of the 5% gelatin phantom samples. To visualize its internal structure, a pie cut was made in the 3D maps. The 3D B-mode scan of the phantom is shown in Fig. 7(a). The preload from the multi-pronged arms caused small displacements at various boundary locations, which can be observed on the phantom's surface in Fig. 7(a). The wave fields for each individual frequency were extracted from the multi-frequency reverberant shear wave field using bandpass frequency filters, each with a 10 Hz bandwidth centered at the corresponding excitation frequency. Figures 7(b) through 7(f) illustrate the 3D reverberant shear wave fields at frequencies of 500 Hz, 1000 Hz, 1500 Hz, 2000 Hz, and 2500 Hz, respectively. The wavelength in the wave fields decreases as the frequency increases. The shear wave propagation originating from the multi-pronged arms is particularly evident. For a more detailed visualization of each extracted wave field, a 2D cross-section in the *xy*-plane at a depth of 0.3 mm is shown in Supplement 1.

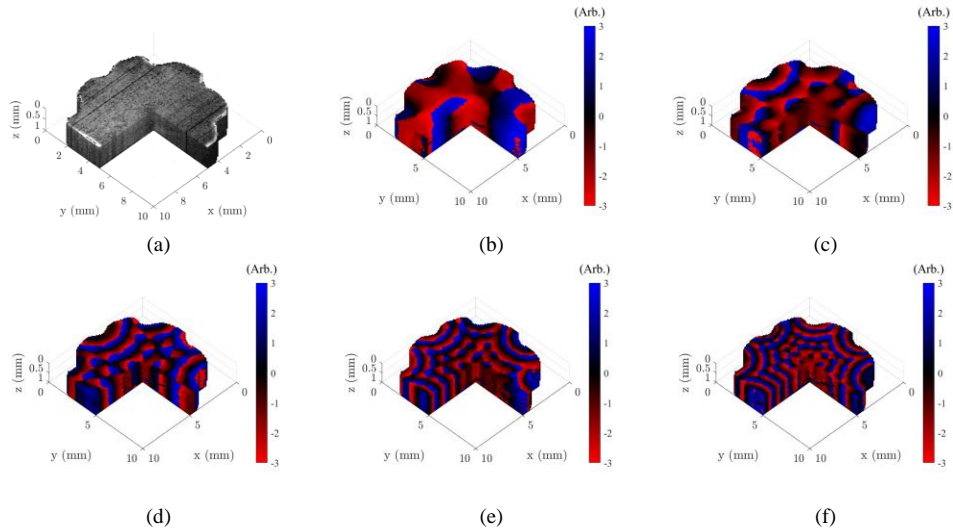

Fig. 7. A 3D map with a pie-cut view to visualize the internal structure of (a) B-mode scan and extracted reverberant shear wave field at: (b) 500 Hz, (c) 1000 Hz, (d) 1500 Hz, (e) 2000 Hz, and (f) 2500 Hz for a 5% gelatin phantom.

The SWS map for each frequency was estimated using the AIA approach. For 500 Hz and 1000 Hz, an autocorrelation window size of 7.9 mm × 7.9 mm was used, whereas a smaller window size of 4.9 mm × 4.9 mm was applied for higher frequencies (1500 Hz, 2000 Hz, and 2500 Hz). A larger autocorrelation window was generally used to ensure uniform SWS estimates, particularly for lower frequencies, due to their longer wavelengths. Figure 8 illustrates the estimated SWS at different frequencies, where a slight increase in SWS with increasing frequency was observed, indicating low dispersion behavior in the phantom.

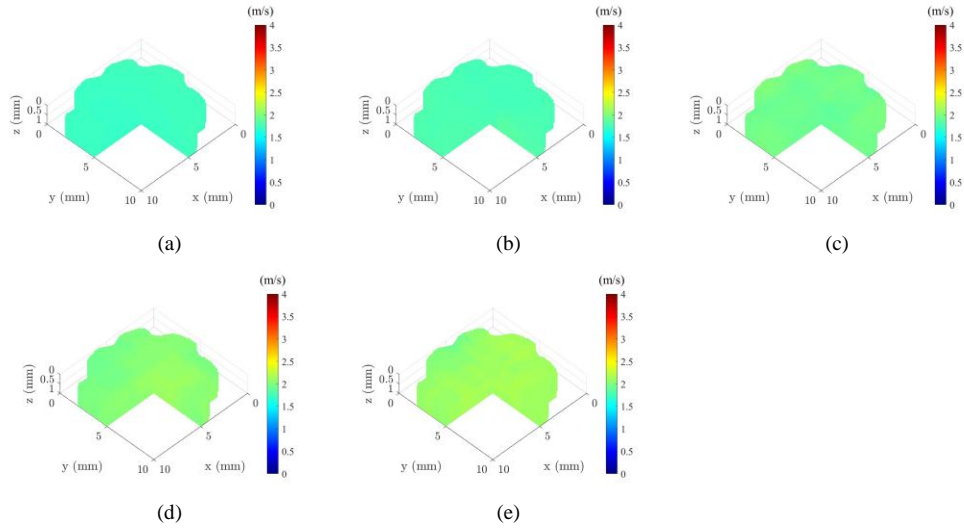

Fig. 8. The 3D SWS map estimated using MFR-OCE on a 5% gelatin phantom across different excitation frequencies: (a) 500 Hz, (b) 1000 Hz, (c) 1500 Hz, (d) 2000 Hz, and (e) 2500 Hz.

The plot in Fig. 9 displays the mean SWS as a function of frequency (black data points), along with a power law fitting (black trendline). The power law fit of SWS has an exponent of 0.13, indicating that the phantom exhibits non-purely elastic behavior with a viscosity component. This viscosity is likely attributed to the 3% concentration of intralipid powder added to the phantom for optical scattering purposes. Additionally, the mean SWS estimated from the reverberant shear wave field in the phantom at three single-frequency experiments (i.e., 1000 Hz, 1500 Hz, and 2000 Hz) is presented in Fig. 9 (orange data points). The power law exponent for these single-frequency excitations was found to be 0.12. The close agreement between the mean SWS values from the multi-frequency and single-frequency experiments across all three frequencies, with a less than 3% difference, validates the accuracy of shear wave field extraction in the MFR-OCE experiment.

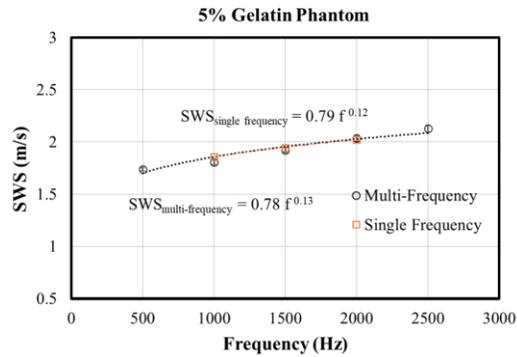

Fig. 9. The mean SWS estimated using MFR-OCE and single-frequency reverberant shear wave OCE on a 5% gelatin phantom across different excitation frequencies along with a power law fits (black and orange trendlines).

## 5.3 MFR-OCE on ex vivo porcine cornea

Figure 10(a) presents the 3D B-mode scan of one of the *ex vivo* porcine cornea samples with a pie cut to reveal its interior. By applying bandpass frequency filters with a bandwidth of 10 Hz, the wave fields for each excitation frequency were estimated. Figures 10(b) through 10(f) show the 3D reverberant shear wave fields at frequencies of 500 Hz, 1000 Hz, 1500 Hz, 2000 Hz, and 2500 Hz, respectively. Generally, the wavelength decreases as the frequency increases.

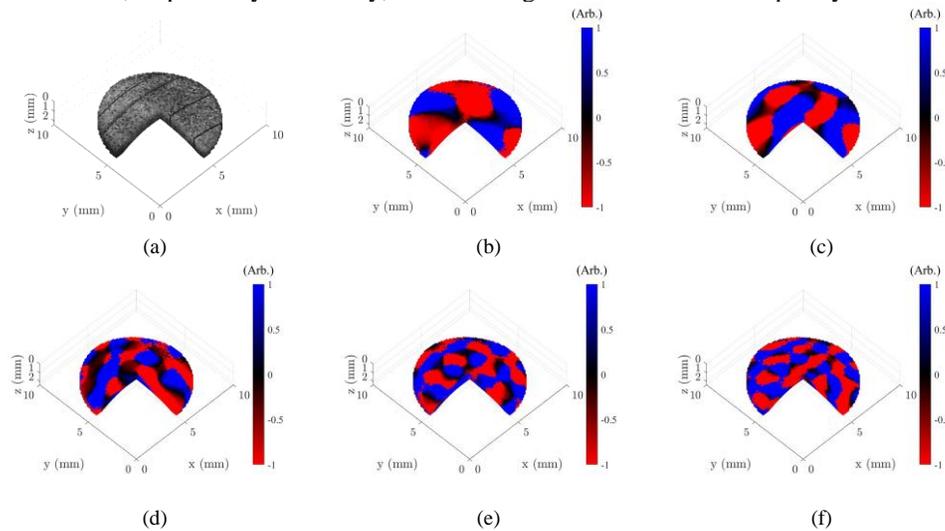

Fig. 10. A 3D map with a pie-cut view to visualize the internal structure of (a) B-mode scan and extracted reverberant shear wave field at: (b) 500 Hz, (c) 1000 Hz, (d) 1500 Hz, (e) 2000 Hz, and (f) 2500 Hz for an *ex vivo* porcine cornea sample.

Figure 11 presents 3D SWS maps estimated using MFR-OCE on an *ex vivo* porcine cornea sample across different excitation frequencies ranging from 500 Hz to 2500 Hz. The results demonstrate a clear frequency-dependent increase in SWS, as indicated by the progressive color shift from blue (low SWS) to yellow/red (high SWS) at higher frequencies. This trend suggests that the cornea exhibits dispersion behavior, where shear wave speed increases with frequency, reflecting its biomechanical properties. The SWS distribution appears relatively uniform in the central region of the cornea, with slightly lower values near the periphery, potentially due to variations in tissue structure or boundary effects. At higher frequencies (2000 Hz and 2500 Hz), the increase in SWS becomes more pronounced, indicating a stiffening response with frequency.

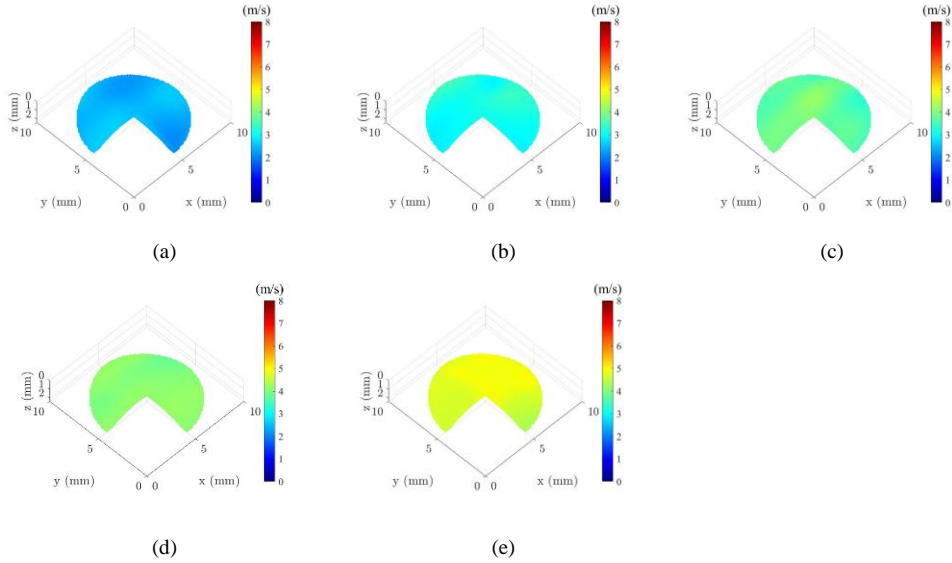

Fig. 11. The 3D SWS map estimated using MFR-OCE on an *ex vivo* porcine cornea sample across different excitation frequencies: (a) 500 Hz, (b) 1000 Hz, (c) 1500 Hz, (d) 2000 Hz, and (e) 2500 Hz.

The mean SWS of the porcine cornea, estimated using MFR-OCE, across the excitation frequencies is illustrated in Fig. 12 (data points). A power law equation with an exponent of 0.33 is fitted with these values (black trendline). The excellent fit demonstrates that the dispersion behavior of the cornea can be accurately described by the power law model. The power law exponent of 0.33 indicates that the cornea exhibits moderate viscoelastic behavior, meaning that higher-frequency shear waves propagate faster due to the tissue's viscoelastic properties. The error bars represent the variability in the measurements, but the data points align well with the fitted power law model.

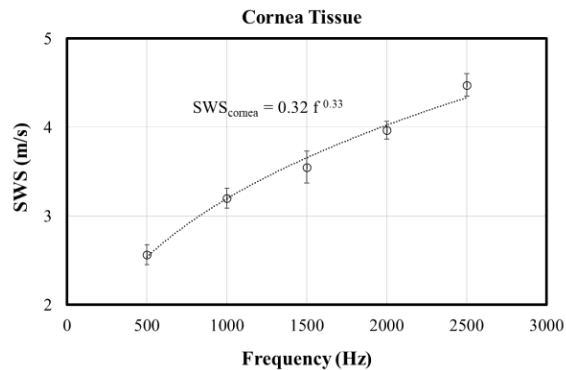

Fig. 12. The mean SWS estimated using MFR-OCE for the porcine cornea samples across different excitation frequencies along with a power law fit (black trendline).

### *5.4 MFR-OCE on ex vivo bovine liver*

The 3D B-mode scan, shear wave fields, and 3D elastography maps of one of the *ex vivo* bovine liver samples across different frequencies are presented in Supplement 1. Figure 13 displays the mean SWS as a function of frequency (data points), along with a power law fitting (black trendline) for *ex vivo* bovine liver. Here again, we have an excellent fit of the power law with the estimated SWS across different frequencies, which shows the viscoelastic behavior of liver tissue can be perfectly defined under the power law equation even in the high-frequency range. The power law fit of SWS for the bovine liver has the exponent of 0.51, indicating a high viscosity component.

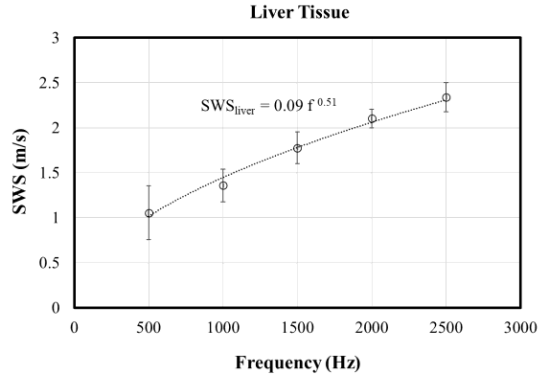

Fig. 13. The mean SWS estimated using multi-frequency OCE on an *ex vivo* bovine liver across different excitation frequencies, along with a power law fit (black trendline)

## 6. Conclusion

This study demonstrated the effectiveness of multi-frequency reverberant shear wave elastography in both simulated and experimental settings, highlighting its potential for precise SWS estimation in heterogeneous media. By generating a reverberant shear wave field and analyzing the wave fields using the AIA approach, we were able to accurately measure SWS across multiple frequencies in a simulated two-sided medium, homogeneous gelatin phantoms, *ex vivo* porcine corneas and an *ex vivo* bovine liver tissues. The results showed that the estimated SWS values closely aligned with the defined model parameters in simulations, with less than 4% error across all frequencies. In the phantom assessment, consistent SWS and power law exponent estimation were observed for both MFR-OCE and single-frequency OCE across three different frequencies. The estimated power law exponent of 0.13 in MFR-OCE, indicates the low dispersion, nearly elastic behavior of the gelatin phantom. Furthermore, there was close agreement between SWS estimates from single-frequency and multi-frequency excitations, with a less than 3% difference, demonstrating that the MFR-OCE approach effectively isolates individual frequency components and provides accurate SWS estimates. The MFR-OCE results for the porcine cornea and bovine liver samples indicated that MFR-OCE is an effective approach to evaluating the viscoelastic behavior of different tissues. The excellent fit of the power law model with the estimated results confirmed that the dispersion behavior of both the cornea and liver can be well-defined by the power law model, even in the high-frequency range. The power law exponent of 0.33 for the porcine cornea and 0.51 for the bovine liver indicated moderate viscoelastic behavior for the cornea and high viscoelastic behavior for the liver. These findings underscore the potential of MFR-OCE as a robust tool for tissue characterization, offering enhanced diagnostic capabilities, particularly in assessing complex viscoelastic materials. Further research is required to address the limitations on the bandwidth, and spatial-temporal resolution of this approach. We utilized a range of 500-2500 Hz shear wave excitations. Lower frequencies have the disadvantage of longer wavelengths requiring larger estimation windows, degrading spatial resolution. Higher frequencies have the disadvantage of higher attenuation and higher temporal sampling requirements. The optimal range may vary with the type of sample under study and will need to be determined experimentally. The method developed here paves the way for future applications in clinical settings, where accurate and detailed tissue elasticity and viscoelasticity measurements are crucial for diagnosis and treatment planning.

**Funding.** NIH (Grant Number: R21AG070331), NIH NEI (Grant Number: P30EY001319), University of Rochester Center of Excellence in Data Science for Empire State Development (Grant Number: 2089A015)

**Disclosures.** The authors declare no conflicts of interest.

**Data availability.** Data underlying the results presented in this paper are available upon request.

**Supplemental document.** See Supplement 1 for supporting content.